\documentclass[aip,apl,reprint,amsmath,amssymb,floatfix]{revtex4-1}
\usepackage{graphicx}

\begin{document}

\title{Transport in organic semiconductors in large electric fields:  From thermal activation to field emission}
\author{J. H. Worne$^{1}$, J. E. Anthony$^{2}$, D. Natelson$^{1,3}$}

\affiliation{Department of Electrical and Computer Engineering, Rice University, 6100 Main St., Houston, TX 77005}
\affiliation{Department of Chemistry, University of Kentucky, Lexington, KY 40506-0055}
\affiliation{Department of Physics and Astronomy, Rice University, 6100 Main St., Houston, TX 77005}

\begin{abstract}
Understanding charge transport in organic semiconductors in the presence of large electric fields is relevant to many potential applications.  Here we present transport measurements in organic field-effect transistors based on poly(3-hexylthiophene) (P3HT) and 6,13-bis(triisopropyl-silylethynyl) pentacene (TIPS-pentacene) with short channels, from room temperature down to 4.2~K.  Near 300~K transport in both systems is well described by a model of thermally assisted hopping with Poole-Frenkel-like electric field enhancement of the mobility.  At low temperatures and large gate voltages, transport in both materials becomes nearly temperature independent, crossing over into a regime described by field-driven tunneling.  These data, particularly in TIPS-pentacene, show that great caution must be exercised when considering more exotic models of low-$T$ transport in these materials.
\end{abstract}

\maketitle



Charge transport in organic semiconductors in
the presence of large electric fields is important in 
organic light-emitting diodes (OLEDs), the channels of 
organic field-effect transistors (OFETs), and organic
photovoltaic devices.   
Much recent work has focused on the effects of
large electric fields on the injection process
\cite{Scott2003,Ng2007,Scheinert2009} as well as
on transport within the semiconductor
bulk\cite{Blom1997,Hamadani2004,Pasveer2005,Limketkai2007,Hamadani2007,Ng2007}.
In general, this is a complex, nonequilibrium problem, involving
charge carriers with moderately strong couplings to vibrational
degrees of freedom, and a disordered environment resulting in
localization and a strongly energy dependent density of states.  As a
result, transport characteristics, usually parametrized by a mobility,
$\mu$, depend nontrivially on temperature, electric field, and carrier
density\cite{Tanase2003,Pasveer2005,Limketkai2007,Tessler2009}.

Near room temperature, the field dependence of the mobility is often
reasonably described (over a limited temperature and field range) by
an effective Poole-Frenkel (PF) model of mobility\cite{Frenkel1938}.
In this model\cite{Frenkel1938}, mobility
$\mu_{\mathrm{PF}} \propto \mu_0(T)
\exp(\gamma \sqrt{E})$, where $\mu_0$ is the zero-field mobility and $E$ is the electric field. Both
$\mu_0$ and $\gamma$ vary like $1/T$ in this limited
range\cite{Blom1997,Hamadani2004,Hamadani2007}.  In the Poole-Frenkel
regime, one can think of the field dependence arising from
field-induced distortion of the disorder potential of the localized
carrier states.

At lower temperatures and strong source-drain fields, it has been
noted\cite{Hamadani2004} that conduction in OFETs is highly nonlinear
and approaches a temperature-independent regime as $T \rightarrow 0$.
The physical mechanism for this temperature independence is 
in dispute.  Recent work done by Dhoot, \textit{et al.} \cite{Dhoot2006} 
argued that the crossover to temperature-independent nonlinear
conduction at low temperatures and large gate voltages was a
signature of a voltage-driven insulator-to-metal transition.
Prigodin and Epstein\cite{Prigodin2007} argue instead that what
is observed is a field-induced crossover from thermal activation to
a field emission hopping regime\cite{Larkin1982} previously examined by Shklovskii\cite{Shklovskii1973}.  This point of view is
extended by Wei~\textit{et al.}\cite{Wei2009}, who find that a multistep
tunneling model explains the observed strong dependences on gate
and source-drain voltage.  More recently, 
Yuen \textit{et al.} \cite{Yuen2009} have looked at the evolution of
carrier transport from 300K to 4K and below, and claim further
evidence for an insulator-to-metal transition in devices
based on the 
and poly(2,5-bis(3-tetradecylthiophen-2-yl)thieno[3,2-b]thiophene)
(PBTTT).  Specifically, they contend that their data are
well described by a Tomonaga-Luttinger Liquid (TLL) model of 
transport originally developed for truly one-dimensional metals.

To address this controversial issue, we performed transport
experiments from room temperature down to cryogenic temperatures in
short-channel bottom-contact OFET devices based on two different
molecules, P3HT \cite{McCullough1993} and 6,13-bis(triisopropyl-silylethynyl)
(TIPS)-pentacene \cite{Park2007}.  P3HT is a polymer semiconductor
that tends toward glassy or nanocrystalline structure, while
TIPS-pentacene is a solution-processable small molecule that forms van
der Waals bonded molecular crystals.  As the temperature is decreased,
we observe that charge carrier behavior evolves from
Poole-Frenkel-like activated hopping at high temperatures to
temperature-independent hopping consistent with field emission at low
temperatures in both P3HT and TIPS-pentacene systems.  While the TLL
analysis approach\cite{Yuen2009,Aleshin2006} produces compelling
plots, we argue that this is fortuitous, particularly since there is
no reason to expect TLL physics to be relevant in the TIPS-pentacene
case.

We used degenerately doped p-type silicon with 200~nm of thermally
grown oxide, which serves both as the substrate and the gate in our
experiments. Platinum electrodes were fabricated using standard
electron beam lithography, electron beam evaporation and liftoff
processing with channel widths, $W$, of 50$\mu$m and channel length,
$L$, of 300nm (device A) and $W = 200\mu$m and $L = 1\mu$m (device B).
On both substrates other electrodes of fixed width and varying channel
lengths were prepared for transmission line estimates of the contact
resistance.  Samples were rinsed using isopropanol and acetone
followed by an oxygen plasma cleaning for two minutes. Samples were
then spin-coated with hexamethyldisilazane (HMDS) at 3000 RPM for 30
seconds, followed by a bake at 130 ºC for 20 minutes.

\begin{figure}[ht]
\begin{center}
\includegraphics[clip, width=8cm]{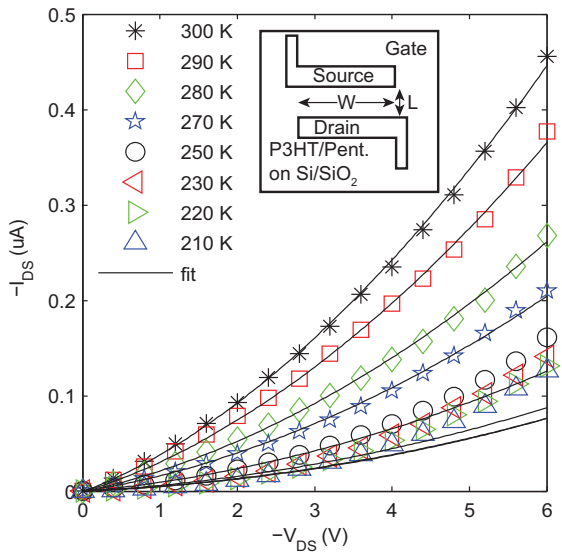}
\caption{$I_{\mathrm{D}}-V_{\mathrm{DS}}$ curves for device A over a 100~K temperature range at $V_g = -80V$.  Fit lines are generated from a Poole-Frenkel-like field dependence of the mobility, as explained in the text. The deviation from theory as T decreases indicates the beginning of the crossover from activated hopping into field emission.}
\label{fig:fig1}
\end{center}
\end{figure}

P3HT was spin-cast from chloroform at a 0.1\% by weight concentration
onto device A and TIPS-pentacene was drop cast from toluene at a 1\%
by weight concentration onto device B. Samples were measured in a
variable temperature probe station with base pressure of ~1x10$^{-6}$
Torr.  Transmission line measurements on the cofabricated device
arrays were made on chip to measure device mobilities, which were
4.6x10$^{-2}$ cm$^2$/Vs (P3HT) and 1.1x10$^{-4}$ cm$^2$/Vs
(TIPS-pentacene), respectively.  We then extracted contact resistances
for our devices and determined that they are much smaller than the
resistance within the device channel, indicating these devices are
bulk dominated.  As shown in previous work\cite{HamadaniAPL}, as the
temperature is reduced, bulk channel resistance increases more rapidly
than the contact resistance, so that our measurements remain bulk
limited down to low temperatures.  Measurement source-drain voltages
$V_{\mathrm{DS}}$ were set so that the average source-drain electric
fields in the channels of both device A and B were similar, with a
maximum electric field of 20~MV/m in device A and 10~MV/m in device B.

We confirmed that our P3HT sample follows PF behavior at high $T$, as
shown in figure \ref{fig:fig1}. The inset of figure \ref{fig:fig1} is
a cartoon describing our device geometry.  The data shown here are 
at one particular gate voltage, $V_{\mathrm{G}}$, at a variety of
temperatures.  At each temperature data for all gate voltages are
analyzed\cite{Hamadani2007} using the form
\begin{equation}
I_{\mathrm{D}}=\frac{\mu_{0}wC_{i}}{L}\exp(\gamma \sqrt{V_{\mathrm{DS}}/L}) \left[(V_{\mathrm{G}}-V_{\mathrm{T}})V_{\mathrm{DS}}- \frac{V_{\mathrm{DS}}^{2}}{2}\right],
\label{eq:PF}
\end{equation}
where $\mu_{0}$ is the (gate- and temperature-dependent) zero-field
mobility, $C_{i}$ is the capacitance per area of the gate oxide,
$\gamma$ is a prefactor that is found to vary like $1/T$, and
$V_{\mathrm{T}}$ is the threshold voltage.  As $T$ decreases, we note
a deviation of the data from the PF model. This will be addressed
below.  Analysis of the TIPS-pentacene data is qualitatively 
identical.

We tried plotting our data in the manner suggested by the TLL analysis\cite{Yuen2009,Aleshin2006}.  The expression for the current in the TLL picture
is\cite{Balents1999,Bockrath1999,Yuen2009}
\begin{equation}
I = I_{0}T^{\alpha+1}\sinh(\gamma' eV/k_{\mathrm{B}}T)|\Gamma((1 + \beta)/2 + i V/\pi k_{\mathrm{B}}T)|^{2},
\label{eq:TLL}
\end{equation}
where $\Gamma$ is the Gamma function, $\alpha$ and $\beta$ are
phenomenological exponents estimated from plots of conductance vs. $T$
and $I(V)$, respectively, and $\gamma'$ is a phenomenological 
parameter thought to be related to the amount of disorder (tunneling
barriers) along the 1d structure.  Based on this equation, the idea is (at
fixed $V_{\mathrm{G}}$) to plot $I_{\mathrm{D}}/T^{\alpha+1}$
vs. $(eV_{\mathrm{DS}}/k_{\mathrm{B}}T)$, where $\alpha$ is a fit
parameter.  If a particular value of $\alpha$ collapses all of the
data over the whole temperature range onto a single curve, it is
tempting to conclude that the TLL picture is valid.  Critical to the
TLL theory validity is that the system in question be one-dimensional.
In polymers such as PBTTT and P3HT, carrier mobility along the polymer
chain is generally much higher than between chains, implying that they
may be relevant.

\begin{figure}[ht]
\begin{center}
\includegraphics[clip,width=8cm]{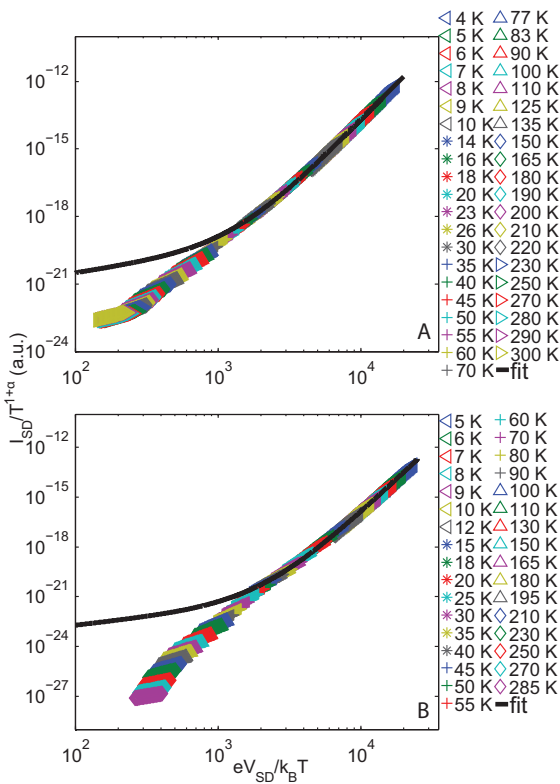}
\caption{Plotting $I_{\mathrm{D}}$ vs. $V_{\mathrm{DS}}$ in scaled coordinates as suggested by Eq.~(\ref{eq:TLL}).  Top data is on device A (P3HT,  $V_{\mathrm{G}} = -80$~V), while bottom data is on device B (TIPS-pentacene,  $V_{\mathrm{G}} = -70$~V).  Solid lines are fits to Eq.~(\ref{eq:TLL}).  For device A, $\alpha = 5.43$, $\gamma'=4\times 10^{-3}$; for device B, $\alpha=7.1$, $\gamma'=3\times 10^{-3}$; for both fits, we used the theoretical expectation $\beta = \alpha+1$.  As explained in the text, the apparent scaling collapse is fortuitous, rather than the result of Tomonaga-Luttinger Liquid physics.}
\label{fig:fig2}
\end{center}
\end{figure}

We present the same style of analysis in Figure~\ref{fig:fig2}, with
the P3HT data shown in A and the TIPS-pentacene data shown in B.  We
are able to collapse our data onto a single line as $T$ is decreased,
with choices for $\alpha$ (5.43 for P3HT, 7.1 for TIPS-pentacene) that
are not wildly different from those reported\cite{Yuen2009} for PBTTT
(5.4, 4.3) or polyaniline fibers\cite{Aleshin2006} (5.5).  Recall that
TIPS-pentacene is a short chain molecule, without the mobility
anisotropy found in PBTTT or P3HT. It seems extremely unlikely that
TIPS-pentacene can be described by TLL theory for a one-dimensional
metal, despite the apparent collapse of its
$I_{\mathrm{D}}-V_{\mathrm{DS}}$ curves onto a single master line.

With the freedom to adjust $\alpha$, plotting scaled data as in
Fig.~\ref{fig:fig2} becomes unwise.  As $T$ decreases, both P3HT and
TIPS-pentacene $I_{\mathrm{D}}-V_{\mathrm{DS}}$ curves become
increasingly non-linear and temperature independent. Plotting the data
of Figure \ref{fig:fig2} turns roughly power law trends over a limited
voltage range into a linear segment on such a log-log plot.  The
freedom to choose $\alpha$ while plotting allows fine-tuning of the
subsequent temperature curves to lie on the same line.  Because
current and voltage are plotted as $I_{\mathrm{D}}/T^{\alpha+1}$ and
$eV_{\mathrm{DS}}/k_{\mathrm{B}}T$, decreasing $T$ moves subsequent
temperature data sets up and to the right on the graph, \textit{even
  if the data themselves do not change with temperature at all.}
Data collapse with this plotting procedure is not sufficient to 
demonstrate TLL physics.

\begin{figure}[ht]
\begin{center}
\includegraphics[clip,width=8cm]{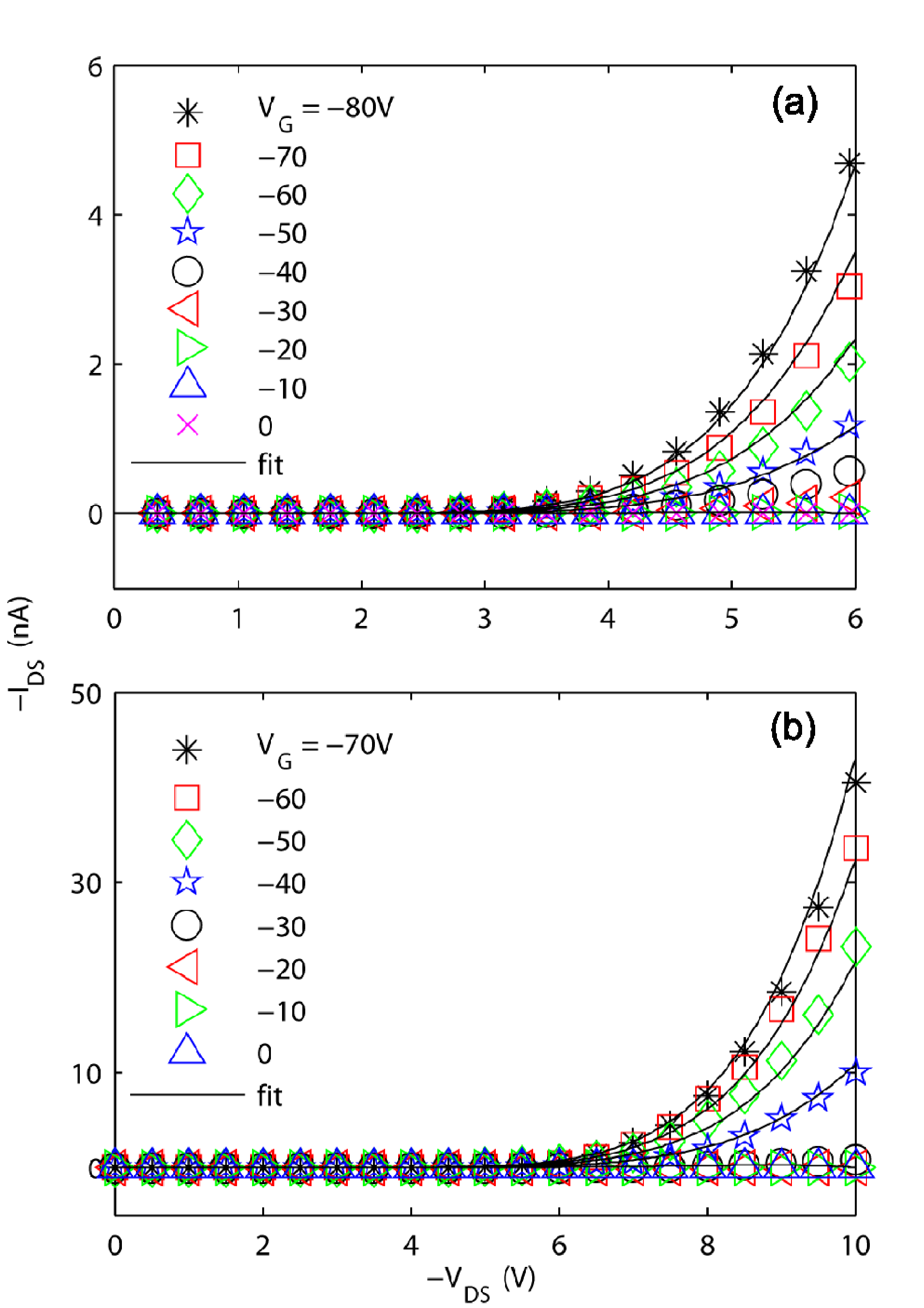}
\caption{Top is P3HT (device A) measured at various gate voltages at 4.2~K; bottom is TIPS-pentacene (device B) measured at 5K.  Black lines are fits using the field emission hopping model expression for mobility, $\mu \propto \exp(-(E_{0}/E)^{1/2})$.}
\label{fig:fig3}
\end{center}
\end{figure}

Instead, as $T$ decreases, we propose that carrier transport evolves
from activation hopping into field emission
hopping\cite{Prigodin2007,Wei2009}, where $\mu_{PF}$ becomes $\mu_{FE}
\propto \mu_0 \exp(-\sqrt{E_{0}/E})$ in an equation analogous to
Eq.~(\ref{eq:PF}).  Here $E=V_{\mathrm{DS}}/L$ is the average electric
field within the channel, while $E_0(V_{\mathrm{G}})$ is temperature
independent and is expected\cite{Prigodin2007} to depend on the
disorder of the sample.  We plot the data for our lowest temperature
$I_{\mathrm{D}}-V_{\mathrm{DS}}$ curves in figure \ref{fig:fig3} and
note that near this temperature our data are well fit by a
temperature-independent $\mu_{0}$.  A detailed comparison to the
multiple tunneling model\cite{Wei2009} will require further extensive
data, particularly examining the effects of gate voltage on 
$E_0$ and attempting to analyze the crossover regime of temperature
and voltage.

In conclusion, we present data that illustrates both the high and low
temperature transport behavior of two chemically unique organic
semiconductors.  The data are consistent with a crossover from
Poole-Frenkel-like activated hopping near room temperature to a
temperature-independent field emission hopping process ($\sim
\exp(-\sqrt{E_{0}/E})$) at cryogenic temperatures.  We further find that a
scaling approach is insufficient to test for Tomonaga-Luttinger Liquid
physics, as it shows apparent TLL consistency even in TIPS-pentacene,
a material that has no microscopic basis for TLL physics.  Further
detailed investigations should be able to test sophisticated models
for the full temperature and voltage dependence of transport in these
rich material systems.

\end{document}